\documentclass[aps,floats,twocolumn,prd,nofootinbib,superscriptaddress]{revtex4-2}

\usepackage{fnpct} 
\usepackage{comment}
\usepackage{ifpdf}
\usepackage{slashed}
\usepackage{color}
\usepackage[mathscr]{eucal}
\usepackage[utf8]{inputenc}
\usepackage{cancel}
\usepackage{bm}

\usepackage{braket}

\usepackage[most]{tcolorbox}

\tcbset{colback=white, colframe=black, 
        highlight math style= {enhanced, 
            colframe=red,colback=red!10!white,boxsep=0pt}
        }

\ifpdf
  \usepackage{graphicx}     
  \usepackage[bookmarksopen,colorlinks=true,linkcolor=bblue,citecolor=bblue,urlcolor=ppink]{hyperref}
\else    
\fi

\definecolor{red}{rgb}{1,0,0}
\definecolor{darkred}{rgb}{0.6,0,0}
\definecolor{darkgreen}{rgb}{0.992447,0.623778,0.034597}
\definecolor{ppink}{rgb}{1,0.4,0.4}
\definecolor{bblue}{rgb}{0.284602,0.317763,0.963947}
\definecolor{purple}{rgb}{0.5 ,0, 0.7}
\definecolor{MyBlue}{HTML}{144893}
\definecolor{MyBlue2}{HTML}{1985B6}

\definecolor{dgreen}{rgb}{0 ,0.5, 0.5}

\newcommand{\dd}{\mathrm{d}}

\usepackage[normalem]{ulem}
\definecolor{mypink}{HTML}{C34D85}

\def\bx{\boldsymbol{x}}

\def\bv{\boldsymbol{v}}
\newcommand{\hatv}{\hat{\bv}}
\newcommand{\hatx}{\hat{\bx}}

\newcommand{\ddd}{{\rm d}}

\newcommand{\be}{\begin{eqnarray}}

\newcommand{\ee}{\end{eqnarray}}

\makeatletter
\newcommand\footnoteref[1]{\protected@xdef\@thefnmark{\ref{#1}}\@footnotemark}
\makeatother

\allowdisplaybreaks[1]

\begin{document}

\title{
Note on pulsar timing array correlation functions induced by peculiar velocities
}

\author{Neha Anil Kumar}
\email{nanilku1@jhu.edu}
\affiliation{
 William H.\ Miller III Department of Physics \& Astronomy, Johns Hopkins University, 3400 N.\ Charles St., Baltimore, MD 21218, USA
}

\author{Keisuke Inomata}
\email{inomata@jhu.edu}
\affiliation{
 William H.\ Miller III Department of Physics \& Astronomy, Johns Hopkins University, 3400 N.\ Charles St., Baltimore, MD 21218, USA
}

\author{Marc Kamionkowski}
\email{kamion@jhu.edu}
\affiliation{
 William H.\ Miller III Department of Physics \& Astronomy, Johns Hopkins University, 3400 N.\ Charles St., Baltimore, MD 21218, USA
}

\begin{abstract}
Several papers have recently calculated the contribution to pulsar timing array overlap reduction functions (ORFs) induced by our peculiar velocity with respect to the rest frame of the stochastic gravitational-wave background.  Here we show that a harmonic-space calculation confirms the most recent result. 
We note that, with the harmonic-space calculation, the ORFs for spin-1 GWs and the correlations with astrometry measurements are also easily obtained.
\end{abstract}

\maketitle

Evidence for the detection of a nano-Hertz stochastic gravitational-wave background (SGWB) from pulsar-timing arrays (PTAs)~\cite{NANOGrav:2023gor,EPTA:2023fyk,Reardon:2023gzh,Xu:2023wog,Miles:2024seg} has led theorists to consider new observables, beyond the standard Hellings-Downs correlation, to further characterize the background and lead to a better understanding of its origins.  One such possibility is the set of observable consequences of our peculiar velocity with respect to an isotropic SGWB \cite{Tasinato:2023zcg,Cruz:2024svc,Mentasti:2025ywl,Blumke:2025nrq}.  Such a peculiar velocity induces a dipolar power asymmetry at linear order in the peculiar velocity, and a quadrupolar asymmetry at quadratic order.  The timing-residual correlations, or overlap reduction functions (ORFs), induced by these asymmetries have been calculated in several recent papers \cite{Tasinato:2023zcg,Cruz:2024svc,Blumke:2025nrq}.  Although results for the dipole agree, those for the quadrupole presented in Ref.~\cite{Blumke:2025nrq} disagree with those in Refs.~\cite{Tasinato:2023zcg,Cruz:2024svc}.

Here we note that these ORFs are obtained very easily from a recently developed harmonic-space approach \cite{Hotinli:2019tpc,AnilKumar:2023hza,AnilKumar:2023yfw}.  The simplicity of the expressions for the ORFs in this approach provides less room for algebraic, transcription, or coding errors than in the traditional approach.  A quick numerical evaluation verifies the results of the quadrupole expression of Ref.~\cite{Blumke:2025nrq} and not that of Refs.~\cite{Tasinato:2023zcg,Cruz:2024svc}.  The approach also allows us to immediately obtain results for the ORFs that arise at orders higher than quadratic in the peculiar velocity and to generalize also to spin-1 GW backgrounds.  We note that expressions for the correlations with astrometry surveys then follow from Ref.~\cite{Inomata:2024kzr}.

To begin, Refs.~\cite{Cusin:2022cbb,Tasinato:2023zcg,Cruz:2024svc,Blumke:2025nrq} have noted that the SGWB specific intensity $I(f,\hat {\bm n})$ at frequency $f$ in direction $\hat {\bm n}$, in a frame boosted with velocity $\beta$ and direction $\hat{\bm v}$, is related to the isotropic intensity $I'(f)$ in the unboosted frame by
\begin{align}
    \label{eq:I_expand}
    &\frac{I(f,\hat{\bm n})}{I'(f)} = \left[ 1 - \frac{\beta^2}{6}(1-n_I^2-\alpha_I) \right] + \beta(1-n_I)(\hat {\bm n} \cdot \hat {\bm v}) \nonumber \\ 
    & \qquad + \frac{\beta^2}{2}(2- 3 n_I + n_I^2 + \alpha_I) \left( (\hat {\bm n} \cdot \hat {\bm v})^2 - \frac{1}{3} \right) + \mathcal O(\beta^3),
\end{align}
where $n_I(f) = \ddd \ln I'/\dd \ln f$ and $\alpha_I(f) = \ddd n_I/\dd \ln f$.

The overlap reduction function $\Gamma_{ab}$ for two pulsars at locations $\hat {\bm n}_a$ and $\hat {\bm n}_b$ can in this boosted background be written,
\begin{align}
    &\Gamma_{ab} = \left[ 1 - \frac{\beta^2}{6}(1-n_I^2-\alpha_I) \right] \Gamma^{(0)}_{ab} + \beta(1-n_I)\Gamma^{(1)}_{ab} \nonumber \\ 
    & \qquad + \frac{\beta^2}{2}(2- 3 n_I + n_I^2 + \alpha_I) \Gamma^{(2)}_{ab} + \mathcal O(\beta^3),
\end{align}
where $\Gamma_{ab}^{(0)}$ is the usual Hellings-Downs correlation \cite{Hellings:1983fr}, 
and
\begin{align}
    \Gamma_{ab}^{(1)} & = \left[ -\frac{1}{8} - \frac{3}{4}x_{ab} - \frac{3}{4}\frac{x_{ab}}{\left( 1- x_{ab}\right)}\ln x_{ab}\right]   \left[ \hat{\boldsymbol{v}} \cdot\hat{\boldsymbol{x}}_a + \hat{\boldsymbol{v}} \cdot\hat{\boldsymbol{x}}_b \right], \label{eq:dipoleHD}
\end{align}
is the dipole ORF \cite{Mingarelli:2013dsa,Tasinato:2023zcg,Cruz:2024svc,Blumke:2025nrq}, with $x_{ab} \equiv (1 - \cos \theta)/2$, and $\cos \theta=\hat{\boldsymbol{x}}_a \cdot\hat{\boldsymbol{x}}_b$.
The expression for the quadrupole contribution to the ORF is a bit more cumbersome, with Ref.~\cite{Blumke:2025nrq} quoting the following result:
\begin{widetext}
    \begin{eqnarray}
        \Gamma_{ab}^{(2)} &= &\frac{1}{(x_{ab} - 1)^2}\Bigg\{ \left(
        \frac{1}{4}x_{ab}^3 - x_{ab}^2 + \frac{3}{4}x_{ab}\ln x_{ab} + \frac{1}{2}x_{ab} + \frac{1}{4}
        \right)
        \left[ (\hatv \cdot \hatx_a)^2 + (\hatv \cdot \hatx_b)^2 
        \right]
        \nonumber \\ 
        &+ &\left(
        \frac{3}{2}x_{ab}^2\ln x_{ab} - \frac{39}{20}x_{ab}^2 + \frac{12}{5}x_{ab} - \frac{9}{20}
        \right)
        \left[ (\hatv \cdot \hatx_a)(\hatv \cdot \hatx_b)
        \right]
        \nonumber \\ 
        &+ &\left(
        x_{ab}^3\ln x_{ab} - \frac{22}{15}x_{ab}^3 - \frac{1}{2}x_{ab}^2\ln x_{ab} + \frac{35}{12}x_{ab}^2 - \frac{1}{2}x_{ab}\ln x_{ab} - \frac{43}{30}x_{ab} - \frac{1}{60}
        \right)
        \Bigg\}\,.
    \end{eqnarray}
\end{widetext}
This result disagrees with the previously computed expression for the quadrupole contribution derived first in Refs.~\cite{Tasinato:2023zcg,Cruz:2024svc} as:
\begin{eqnarray}
    \bar{\Gamma}_{ab}^{(2)} &= &\frac{3}{4}\left[\frac{13x_{ab} - 3}{10(1 - x_{ab})} + \frac{x_{ab}^2\ln x_{ab}}{(1 - x_{ab})^2}\right]
    \left[ (\hatv \cdot \hatx_a)(\hatv \cdot \hatx_b)
    \right] 
    \nonumber \\ 
    &+ &\left[
    \frac{1 + 2x_{ab}-4x_{ab}^2 + x_{ab}^3 + 3x_{ab}\ln x_{ab}}{8(1 - x_{ab})^2}
    \right]
    \nonumber \\ 
    & &\times \left[ (\hatv \cdot \hatx_a)^2 + (\hatv \cdot \hatx_b)^2 
    \right]\,,
\end{eqnarray}
where the original result from Refs.~\cite{Tasinato:2023zcg,Cruz:2024svc} has been re-normalized to match the conventions used in Ref.~\cite{Blumke:2025nrq} in defining $I(f,\hat{\bm n})$ [Eq.~\eqref{eq:I_expand}], and we have used $\bar{\Gamma}_{ab}^{(2)}$ to represent the result from Refs.~\cite{Tasinato:2023zcg,Cruz:2024svc} to distinguish it from the expression derived in Ref.~\cite{Blumke:2025nrq}. 
Not only did Ref.~\cite{Blumke:2025nrq} note that the expressions disagreed, the derivations of these results are quite long in both works.

We now describe how the same ORFs can more straightforwardly be obtained from the  harmonic-space approach \cite{Hotinli:2019tpc,AnilKumar:2023hza,Inomata:2024kzr}.  
If the angular dependence of the SGWB intensity is expanded (at some frequency, which we suppress for notational economy) as $I(\hat {\bm n})/I_0 = \sum_{LM} c_{LM} Y_{LM}(\hat {\bm n})$, then each $LM$ mode provides a contribution,
\begin{equation}
 \tilde \Gamma^{LM}_{ab}= (-1)^L \sqrt{\pi} \sum_{\ell\ell'} F^{L}_{\ell \ell'}\left\{ Y_\ell(\hat {\bx}_a) \otimes Y_{\ell'}(\hat {\bx}_b)\right\}_{LM},
  \label{eq:orfzz}
\end{equation}
where the sums on $\ell\ell'$ are taken from 2 (or 1, for spin-1 GWs) to infinity, and
\begin{align}
  &\left\{ Y_{\ell}(\hat{\bx}_a) \otimes Y_{\ell'}(\hat{\bx}_b)\right\}_{LM} \nonumber \\ 
  &= \sum^{\ell}_{mm'} \Braket{\ell m \ell' m' |L M} Y_{\ell m}(\hat{\bx}_a)Y_{\ell' m'}(\hat{\bx}_b)
  \label{eq: biposh}
\end{align}
is a bipolar spherical harmonic (BiPoSH)~\cite{Hajian:2003qq,Hajian:2005jh,Joshi:2009mj,Book:2011na}.  Here, $\langle \ell m\ell'm'|LM\rangle$ are Clebsch-Gordan coefficients.  The sum on $m$ is from $-\ell$ to $\ell$, and analogously for $m'$.
Expressions for the coefficient $F^L_{\ell \ell'}$ were derived using the harmonic-space BiPoSH formalism in Refs.~\cite{AnilKumar:2023yfw,Inomata:2024kzr}, for both spin-2 and spin-1 GWs.  
When considering only the intensity (unpolarized) contribution to the correlated pulsar residual response of a spin-2 SGWB (as was considered for the kinematic corrections above), the coefficients are given by the following expression:
\begin{eqnarray}
    F_{\ell\ell'}^L = 2 z_\ell z_{\ell'} \begin{pmatrix}
        \ell & \ell' & L \\
        -2 & 2 & 0
    \end{pmatrix} X_{\ell\ell'}^L\,,
\end{eqnarray}
where
\begin{eqnarray}
    z_{\ell} \equiv (-1)^\ell\sqrt{\frac{4\pi(2\ell + 1)(\ell - 2)!}{(\ell + 2)!}}\,,
\end{eqnarray}
for a spin-2 SGWB, and $X_{\ell\ell'}^L \equiv 1$ if $\ell + \ell' + L =$ even, and 0 otherwise.
Routines to evaluate Clebsch-Gordan coefficients and spherical harmonics are ubiquitous in scientific computing software, and the sum in Eq.~(\ref{eq:orfzz}) converges extremely rapidly.  
These ORF expressions are thus very easy to code up (also available at \href{https://github.com/KeisukeInomata0/pyORFs}{github.com/KeisukeInomata0/pyORFs}) and quickly evaluated.
Therefore, given a map of intensity anisotropy in real space, such as the kinematic anisotropy induced by our peculiar motion with respect to an isotropic SGWB [Eq.~\eqref{eq:I_expand}], this methodology only necessitates the derivation of the relevant $c_{LM}$'s characterizing the intensity map to then translate the real space SGWB intensity correction to its impact on the PTA response.

\begin{figure}
        \centering \includegraphics[width=0.99\columnwidth]{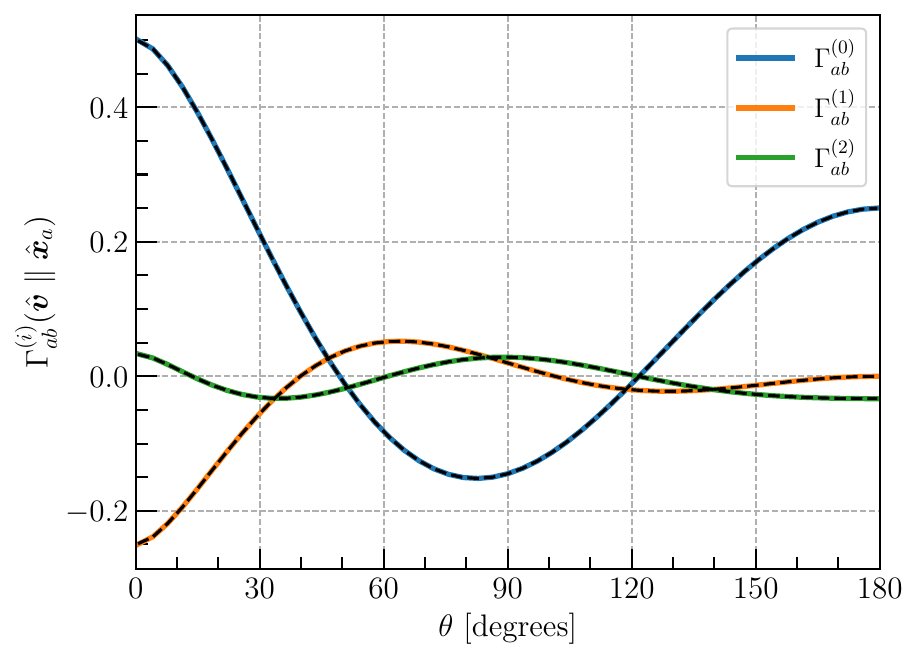}
        \caption{
        Results for the monopole, dipole, and quadrupole ORFs from Ref.~\protect\cite{Blumke:2025nrq} for $\hat {\bx}_a \parallel \hat{\bm v}$, and $\theta = \text{arccos}(\hat {\bm n}_a \cdot \hat {\bm n}_b)$.  The dashed curves are the results from the harmonic-space calculation.}
        \label{fig:compare}
\end{figure}

To demonstrate the power of this compact expression, we now present the ORFs for the PTA response to the kinematic anisotropy in Eq.~\eqref{eq:I_expand} using the technique described above, eventually independently corroborating the result presented in Ref.~\cite{Blumke:2025nrq}.
If the power asymmetries are generated by a peculiar velocity, then the residual symmetry about the direction of the velocity allows us to choose, without loss of generality, $\hat {\bm z}$ to be along the direction $\hat {\bm v}$ of the peculiar velocity.  The only nonzero $c_{LM}$ are thus those for $M=0$.  Including normalizations properly, we find
\begin{align}
    &\Gamma_{ab}^{(0)} = \frac{3\sqrt{4 \pi}}{8\pi} \tilde \Gamma^{00}_{ab},  \quad
       \Gamma_{ab}^{(1)} = \frac{3}{8\pi}\sqrt{\frac{4 \pi}{3}} \tilde \Gamma^{10}_{ab}, \quad \nonumber \\ 
    &\Gamma_{ab}^{(2)} = \frac{3}{8\pi} \frac{2}{3}\sqrt{\frac{4 \pi}{5}} \tilde \Gamma^{20}_{ab}.
\end{align}

\begin{figure}
    \centering
    \includegraphics[width=\linewidth]{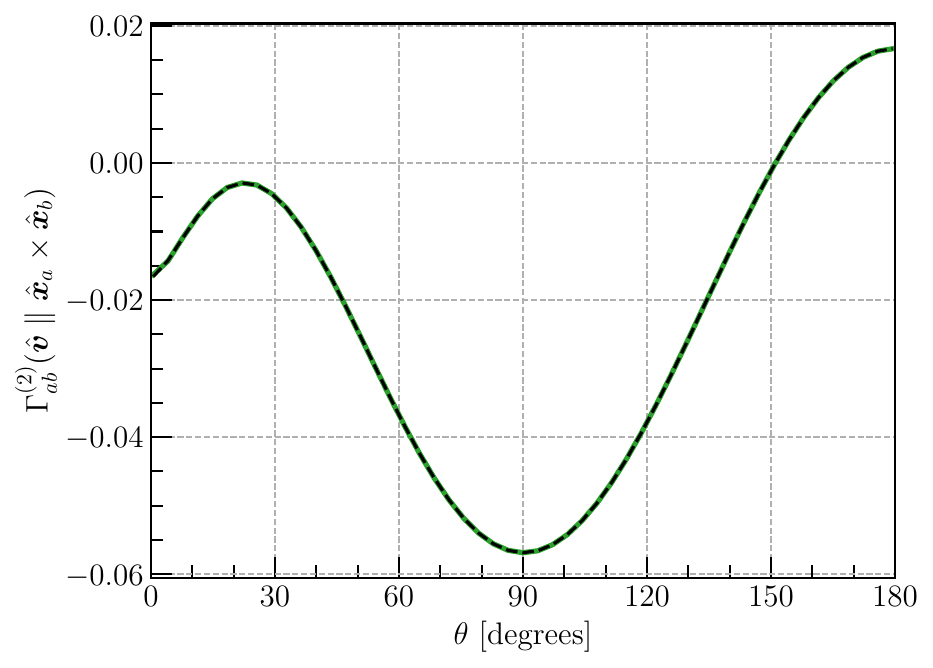}
    \caption{As in Fig.~\ref{fig:compare}, but for $\hat {\bm v} \parallel \hat {\bx}_a \times \hat {\bx}_b$. We show only $\Gamma^{(2)}_{ab}(\hat {\bm v} \parallel \hat {\bm x}_a \times \hat {\bm x}_b)$ because $\Gamma^{(0)}_{ab}(\hat{\bm v } \parallel \hat {\bm x}_a) = \Gamma^{(0)}_{ab}(\hat{\bm v } \parallel \hat {\bm x}_a \times \hat {\bm x}_b)$ in terms of the $\theta$ dependence and $\Gamma^{(1)}_{ab}(\hat{\bm v } \parallel \hat {\bm x}_a \times \hat {\bm x}_b) = 0$. 
    }
    \label{fig:cross_check}
\end{figure}

Numerical evaluation shows that our monopole and dipole expressions agree precisely with prior results.  Our quadrupole results agree precisely with those in Ref.~\cite{Blumke:2025nrq} but not those in Refs.~\cite{Tasinato:2023zcg,Cruz:2024svc}.  (We note that this discrepancy could have also been decided by comparing with quadrupole results in Ref.~\cite{Mingarelli:2013dsa}.)
Given the symmetry about $\hat {\bm v}$, these ORFs most generally depend on three quantities which we can take to be the polar coordinates $\theta_a$ and $\theta_b$ and the relative longitude $\phi_a-\phi_b$.  
Since the full dependence on these three variables is not easily shown graphically, we provide in Figs.~\ref{fig:compare} and \ref{fig:cross_check} results for $\hat {\bm v} \parallel \hat {\bx}_a$ and ${\bm v} \parallel \hat{\bx}_a \times \hat{\bm x}_b$ scenarios, respectively, where the PTA response depends only on the angle between the two pulsars.
We note that the closed form expressions for the ORFs with general $L$ and $M$ in $\hat {\bm v} \parallel \hat{\bm{x}}_a$ are obtained in Ref.~\cite{Gair:2014rwa}, which are consistent with our results~\cite{Inomata:2024kzr}.

We thus weigh in favor of Ref.~\cite{Blumke:2025nrq}.  Although this conclusion was reached by numerical evaluation of the expressions, it should also be possible to check the agreement analytically, generalizing the calculation of Appendix B in Ref.~\cite{Qin:2018yhy}.  This exercise, though straightforward, is likely to be tedious, so we do not include it in this brief note.

As this elementary exercise shows, the simplicity of the harmonic-space expressions leaves far less room for algebraic/coding/transcription errors.  They are also far more general:  The quadrupole result of Ref.~\cite{Blumke:2025nrq} required two dense pages of algebra to derive and then took three widetext lines to display; imagine what the octupole ($L=3$) result would look like!  With the harmonic-space, we can get it (or that for $L=4,5,6,\ldots$) simply by changing one number (i.e., $L$) in the code.

The formalism also allows us to obtain the peculiar-velocity-induced ORFs for a spin-1 GW background by simply using the spin-1 expressions \cite{AnilKumar:2023yfw,Inomata:2024kzr}  for $F^L_{\ell\ell'}$.  Finally, there are prospects for probing a nano-Hertz GW background with astrometry \cite{Braginsky:1989pv,Book:2010pf}, and all the analogous harmonic-space tools for obtaining the effects of peculiar velocity on those correlations \cite{Cruz:2024diu}  are provided in Ref.~\cite{Inomata:2024kzr}.

\medskip
We thank Maximilian Blümke and Marisol Cruz for useful comments on our draft. 
This work was supported by NSF Grant No.\ 2412361,
NASA ATP Grant No.\ 80NSSC24K1226, and the Templeton Foundation.


\bibliography{pta_kinematic_aniso}{}

\end{document}